\documentclass[11pt]{article}

\usepackage{epsf}
\usepackage{latexsym}
\usepackage{amssymb}
\usepackage{amsmath}
\usepackage[dvips]{graphicx}
\usepackage{array}

\def\d3{^{(3)}\nabla}

\begin{document}
\centerline{\Large \bf Cross correlations  from back reaction  on stochastic magnetic fields}
\vskip 2 cm

\centerline{Kerstin E. Kunze
\footnote{E-mail: kkunze@usal.es} }

\vskip 0.3cm

\centerline{{\sl Departamento de F\'\i sica Fundamental} and {\sl IUFFyM},}
\centerline{{\sl Universidad de Salamanca,}}
\centerline{{\sl Plaza de la Merced s/n, 37008 Salamanca, Spain }}

\vskip 1.5cm

\centerline{\bf Abstract}
\vskip 0.5cm
\noindent
The induction equation induces non trivial correlations between the primordial curvature mode and the magnetic mode which is 
a non linear effect.
Assuming a stochastic, gaussian magnetic field the resulting power spectra determining the two point cross correlation functions 
between the primordial curvature perturbation and the magnetic energy density contrast as well as the magnetic anisotropic stress
are calculated approximately. 
The corresponding numerical solutions are used to calculate the angular power spectra determining the temperature anisotropies and 
polarization of the cosmic microwave background, $C_{\ell}$. It is found that the resulting $C_{\ell}$ are sub-leading in comparison to  those generated 
by the compensated mode for a magnetic field which only redshifts with the expansion of the universe.
The main focus are scalar modes, however, vector modes will also be briefly discussed.

\vskip 1cm

\section{Introduction}
\setcounter{equation}{0}
Magnetic fields influence the spectrum of anisotropies of the cosmic microwave background (CMB) by contributing to the initial conditions
as well as the evolution equations of the perturbations.
In the following a Friedmann-Robertson-Walker (FRW) background space-time is assumed with a line element
$ds^2=a^2(\tau)(-d\tau^2+\delta_{ij}dx^idx^j)$, where $a(\tau)$ is the corresponding scale factor. 
Maxwell's equations together with Ohm's law in its simplest form, neglecting any convection term, and assuming 
infinite conductivity leads in a FRW background to \cite{adgr}-\cite{sb}, 
\begin{eqnarray}
\partial_{\tau}\left(a^2\vec{B}\right)=\vec{\nabla}\times\left[\vec{v}_b\times\left(a^2\vec{B}\right)\right],
\label{e1}
\end{eqnarray}
which yields to \cite{bis}
\begin{eqnarray}
\partial_{\tau}\left(a^2\vec{B}\right)=\left(a^2\vec{B}\cdot\vec{\nabla}\right)\vec{v}_b-\left(\vec{v}_b\cdot\vec{\nabla}\right)\left(a^2\vec{B}\right)-a^2\vec{B}\left(\vec{\nabla}\cdot\vec{v}_b\right)
\label{e2}
\end{eqnarray}
where $\vec{v}_b$ is the baryon velocity. Neglecting the righthandside of this equation yields the evolution of the magnetic field 
scaling as $a^{-2}$. This simplifies the perturbation equations as its contribution can be cast in a time-independent way in the mode expansion. This approach has been taken so far in the calculation of the CMB anisotropies \cite{cmbmag1}-\cite{kk2}.
In most works the magnetic field has been assumed to be non helical, except in \cite{kk2}.
As shown first in \cite{helical} and investigated in detail in \cite{kk2} helical magnetic fields cause odd parity modes in the cosmic microwave background (CMB) in terms of non vanishing cross correlation between the polarization E- and B-modes and the temperature and B-mode.

Equation (\ref{e2}) can be linearized by assuming the magnetic field to be of the form $\vec{B}(\vec{x},\tau)=\vec{B}_0+\vec{b}(\vec{x},\tau)$, where 
$\vec{B}_0$ is a background field scaling as $a^{-2}$ and $\vec{b}$ a small perturbation, $|\vec{b}|\ll|\vec{B}_0|$ \cite{kor, kat,jko}. 
Moreover, if in addition it is assumed that 
on the relevant scales  spatial variations of $\vec{B}_0$ can be neglected in comparison to those of $\vec{b}$ then
there is no mode coupling and it is possible to identify the three different types of MHD modes.  These are the two magnetosonic modes and the Alfv\'en mode whose evolution can be studied separately \cite{kat}. Whereas the magnetosonic modes perturb the density as well as the baryon velocity and thus correspond to scalar perturbations the Alfv\'en modes are vector perturbations as the density is left unperturbed. 

In the following the background magnetic field is not assumed to be uniform but rather a gaussian stochastic magnetic field, 
which is assumed to be non helical, for simplicity.
The aim is to investigate the effect of the evolution of the magnetic field due to the baryon velocity.
Assuming the presence of a primordial curvature mode generated during inflation it is shown that there is 
a non vanishing correlation between the magnetic field and the inflationary curvature mode. This is used to 
numerically calculate the effect on the CMB.
The main focus are scalar modes, however, vector modes will also be briefly discussed.

\section{Cross  correlation functions between the primordial curvature and magnetic mode}
\setcounter{equation}{0}

The magnetic field is assumed to be non helical and gaussian so that its
two point function  is assumed to be of the form 
\begin{eqnarray}
\langle B_i^*(\vec{k},\tau)B_j(\vec{k}',\tau')\rangle=\delta_{\tau,\tau'}\delta_{\vec{k}\vec{k}'}P_B(k,\tau)\left(\delta_{ij}-\frac{k_ik_j}{k^2}\right),
\label{2pt}
\end{eqnarray}
where, for simplicity, a zero correlation time is assumed leading to a Markovian stochastic process with a delta function correlation in time.
This is in the line of the quasi-normal approximation which is used in the problem of closure resulting from an infinite hierarchy of 
moment equations in magnetohydrodynamic turbulence \cite{bis}.

\subsection{Scalar modes}
In the longitudinal gauge the fluid velocity coincides with the gauge invariant velocity $V$. Using the expansion in scalar harmonics $Q^{(0)}(\vec{k},\vec{x})$ (e.g. \cite{ks}) and using for the magnetic field
\begin{eqnarray}
a^2 B_i(\vec{x},\tau)=a_0^2\sum_{\vec{k}} B_i(\vec{k},\tau)Q^{(0)}(\vec{k},\vec{x})
\end{eqnarray}
then equation (\ref{e1}) yields to
\begin{eqnarray}
\dot{B}_i(\vec{k},\tau)=\sum_{\vec{q}}\frac{q_iq_j-\vec{k}\cdot\vec{q}\,\delta_{ij}}{q}B_j(\vec{k}-\vec{q},\tau)V_b(\vec{q},\tau).
\label{dotbi}
\end{eqnarray}
The baryon velocity is determined by (e.g. \cite{kk1}, \cite{cmbmag2}-\cite{ls}) 
\begin{eqnarray}
\dot{V}_{\rm b}=(3c_s^2-1){\cal H}V_{\rm b}+k(\Psi-3c_s^2\Phi)+kc_s^2\Delta_{\rm b}+R\tau_c^{-1}(V_{\gamma}-V_{\rm b})+\frac{R}{4}kL,
\label{vb}
\end{eqnarray}
where, in terms of the magnetic energy density $\Delta_B$ and anisotropic stress $\pi_B$ (cf. Appendix),  $L=\Delta_B-\frac{2}{3}\pi_B$ is due to the Lorentz force $\vec{J}\times\vec{B}$ and $R\equiv\frac{4}{3}\frac{\rho_{\gamma}}{\rho_{\rm b}}$. Furthermore, $c_s^2=\frac{\partial\bar{p}}{\partial\bar{\rho}}$ is the adiabatic sound speed and $\tau_c^{-1}$ is the mean free path of photons between scatterings given in terms of the number density of free electrons $n_{\rm e}$ and the Thomson cross section $\sigma_{\rm T}$, $\tau_c^{-1}=an_{\rm e}\sigma_{\rm T}$. Moreover, $\Phi$ and $\Psi$ denote the gauge invariant Bardeen potentials and ${\cal H}=\frac{\dot{a}}{a}$. As can be seen for example in the baryon velocity equation, the magnetic field does not enter linearly, but rather quadratically. 
This is also the case for the remaining perturbation equations (e.g., \cite{cmbmag2}-\cite{kk2}).
Moreover, the equation governing its evolution is an integro-differential equation in the continuum limit. 
This makes an exact treatment of the magnetic field very difficult and therefore here it is assumed that the change in the magnetic field due to the effect of the baryon velocity is small so that
\begin{eqnarray}
B_i(\vec{k},\tau)=B_i^{(0)}(\vec{k})+b_i(\vec{k},\tau),
\label{bi}
\end{eqnarray}
where $|b_i|\ll |B_i^{(0)}|$. The time-independent solution $B_i^{(0)}(\vec{k})$ results in  constant contributions of the magnetic field
to the perturbation equations in terms of its energy density contrast and anisotropic stress. Moreover, these also enter in the initial conditions \cite{cmbmag2}-\cite{kk2}. The linear perturbation equations and initial conditions can be separated in one part proportional to $\Delta_B$ and one proportional to $\pi_{B}$. The final CMB temperature and polarization angular power spectra are determined by summing the two contributions using the auto- and cross correlation functions of  $\Delta_B$ and $\pi_B$ \cite{ls,kk1}.
Using (\ref{bi}) in equation (\ref{dotbi})  the back reaction of the fluid dynamics onto the magnetic field is determined by
\begin{eqnarray}
\dot{b}_i(\vec{k},\tau)=\sum_{\vec{q}}\frac{q_iq_j-\vec{k}\cdot\vec{q}\,\delta_{ij}}{q}B_j^{(0)}(\vec{k}-\vec{q})V_b(\vec{q},\tau).
\label{bdot}
\end{eqnarray}
In the standard $\Lambda$CDM model the CMB temperature anisotropies and polarization are due to the adiabatic mode which is a primordial curvature perturbation, $\zeta(\vec{k})$, generated during inflation.  $\zeta$ is assumed to be a gaussian random variable with two point correlation function 
$\langle\zeta(\vec{k})^* \zeta(\vec{k'})\rangle={\cal P}_{\zeta}(k)\delta_{\vec{k}\vec{k'}}$ and the  power spectrum is defined by ${\cal P}_{\zeta}(k)=\frac{2\pi^2}{k^3}A_s\left(\frac{k}{k_p}\right)^{n_s-1}$. In the numerical solution the best fit values of the 6-parameter  $\Lambda$CDM model of WMAP7 will be used \cite{wmap7}. Taking into account the contribution from a standard adiabatic mode the total baryon velocity is then given by, at zeroth order in the perturbation of the magnetic field,
\begin{eqnarray}
V_b(\vec{k},\tau)=\zeta(\vec{k})V_b^{\zeta}(k,\tau)+\Delta_{B^{(0)}}(\vec{k}) V_b^{\Delta_{B}}(k,\tau)+\pi_{B^{(0)}}(\vec{k})V_{b}^{\pi_B}(k,\tau),
\label{vel}
\end{eqnarray}
where $V_b^{X}$, $X=\zeta,\Delta_{B},\pi_B$, denotes the baryon velocity calculated only considering the contribution $X$.
Therefore the first order perturbation of the magnetic field is determined by
\begin{eqnarray}
b_i(\vec{k},\tau)&=&\sum_{\vec{q}}\frac{q_iq_j-\vec{k}\cdot\vec{q}\,\delta_{ij}}{q}B_j^{(0)}(\vec{k}-\vec{q})
\int_{\tau_i}^{\tau} d\tau'\left[
\zeta(\vec{q})V_b^{\zeta}(q,\tau')+\Delta_{B^{(0)}}(\vec{q}) V_b^{\Delta_{B}}(q,\tau')
\right.
\nonumber\\
&&\left.
\hspace{6.2cm}+\pi_{B^{(0)}}(\vec{q})V_{b}^{\pi_B}(q,\tau')\right]
\label{b1}
\end{eqnarray}
where $\tau_i$ is some initial time and we assume $b_i(\tau_i)=0$.
Therefore at first order the magnetic field contributions, $\Delta_B$ and $\pi_B$, and the primordial curvature perturbation have  non vanishing correlations. 
It is interesting to note that in the tight-coupling limit of photons and baryons, long before last scattering, equation (\ref{bdot}) implies,
\begin{eqnarray}
b_i(\vec{k},\tau)\simeq -\frac{3}{4}\sum_{\vec{q}}\frac{q_i q_j-\vec{k}\cdot\vec{q}\, \delta_{ij}}{q^2}B_j^{(0)}(\vec{k}-\vec{q})\Delta_{\gamma}(\vec{q},\tau),
\end{eqnarray}
using $V_{\gamma}\simeq V_{b}$  and the evolution of the photon density contrast (e.g. \cite{ks}), 
\begin{eqnarray}
\dot{\Delta}_{\gamma}(\vec{k},\tau)=-\frac{4}{3}kV_{\gamma}(\vec{k},\tau).
\end{eqnarray}
Therefore in the tight-coupling regime  the change in the magnetic field is driven by the sum over all modes of the photon density perturbation
weighted by the amplitude of the initial magnetic field.
A similar, but simpler relation between the magnetic field perturbation and density perturbation was found in \cite{kat} for  an overall constant magnetic field 
$\vec{B}^{(0)}$.

\begin{figure}[h!]
\centerline{\epsfxsize=3.2in\epsfbox{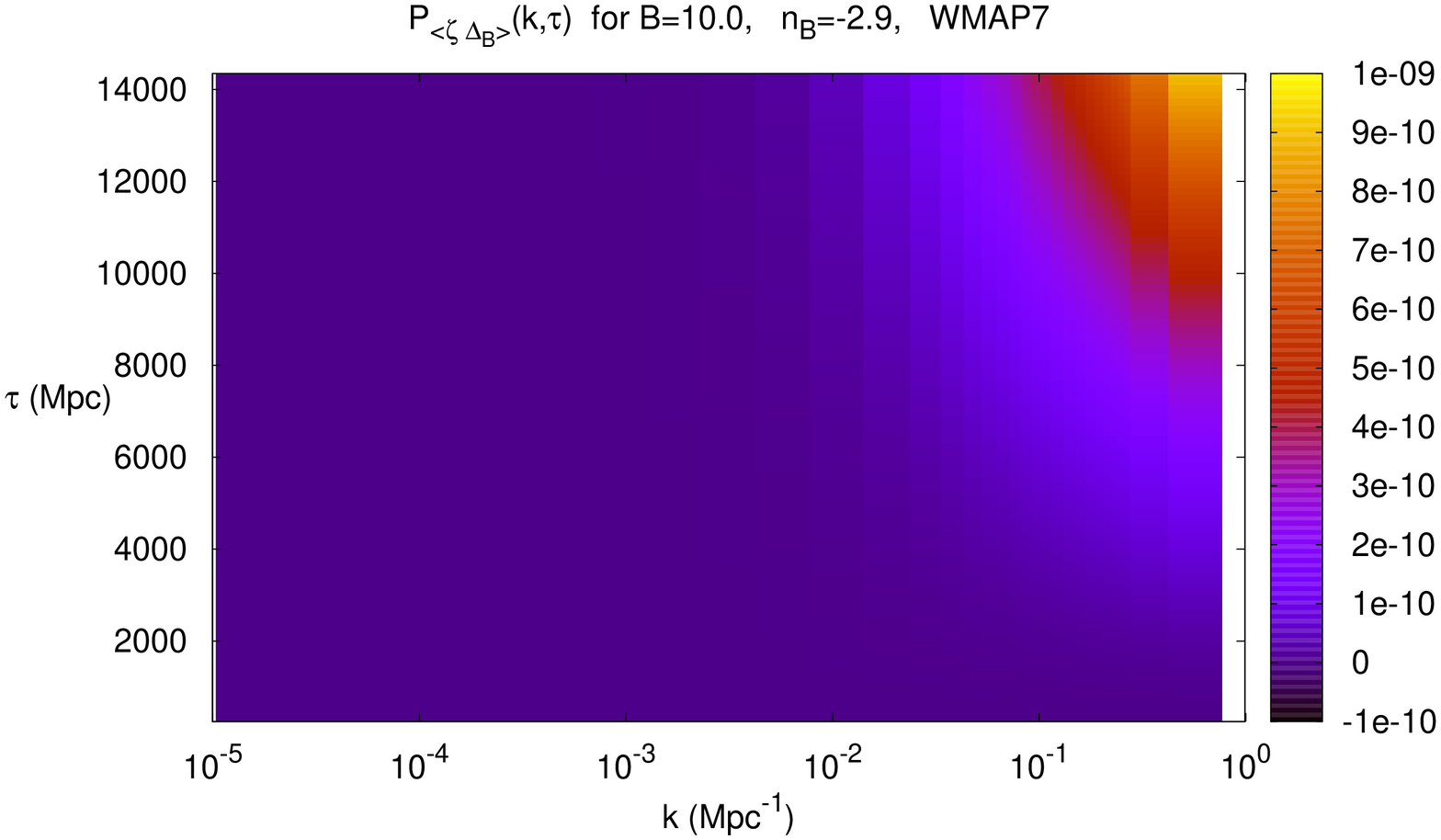}
\hspace{0.2cm}
\epsfxsize=3.2in\epsfbox{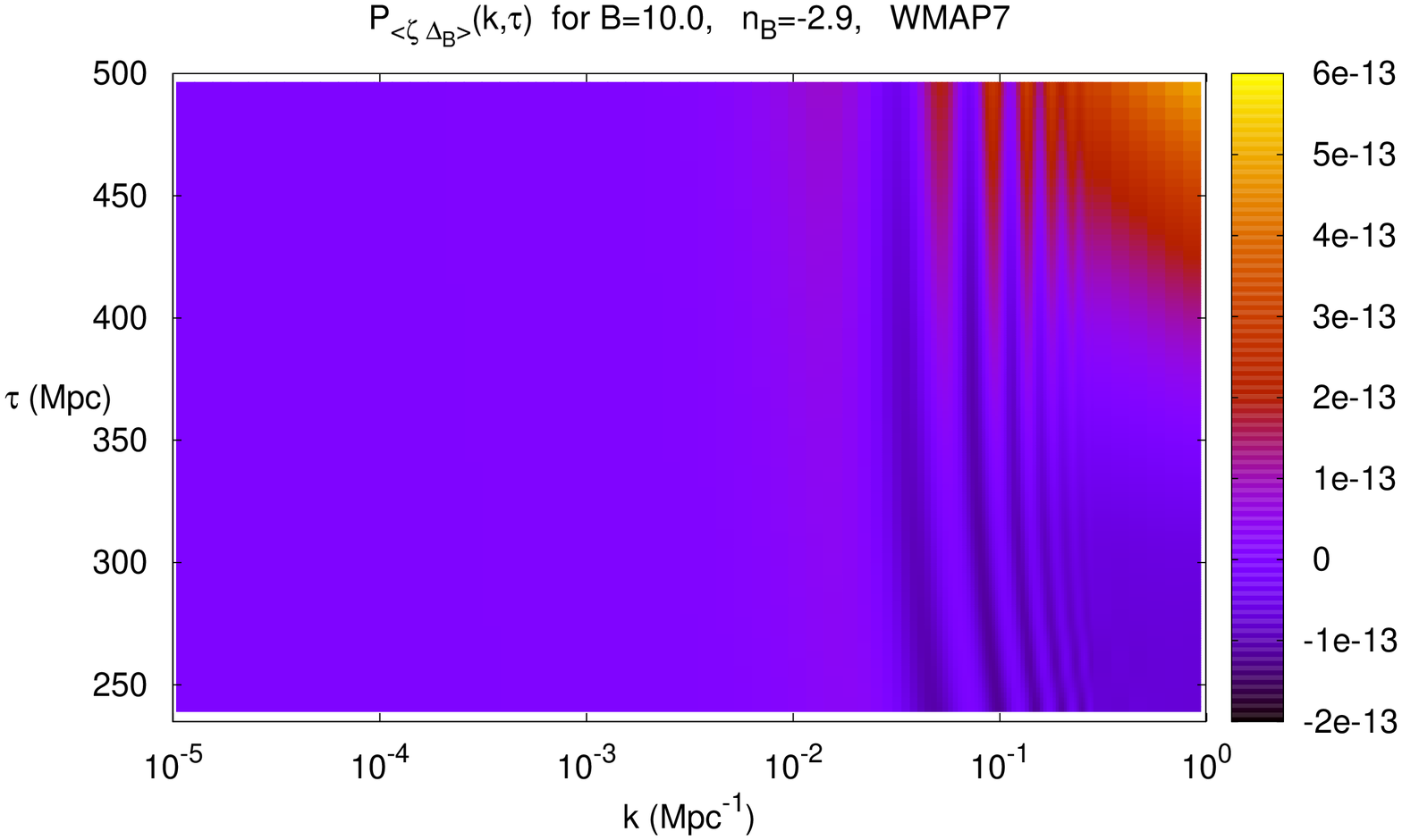}}
\centerline{\epsfxsize=3.2in\epsfbox{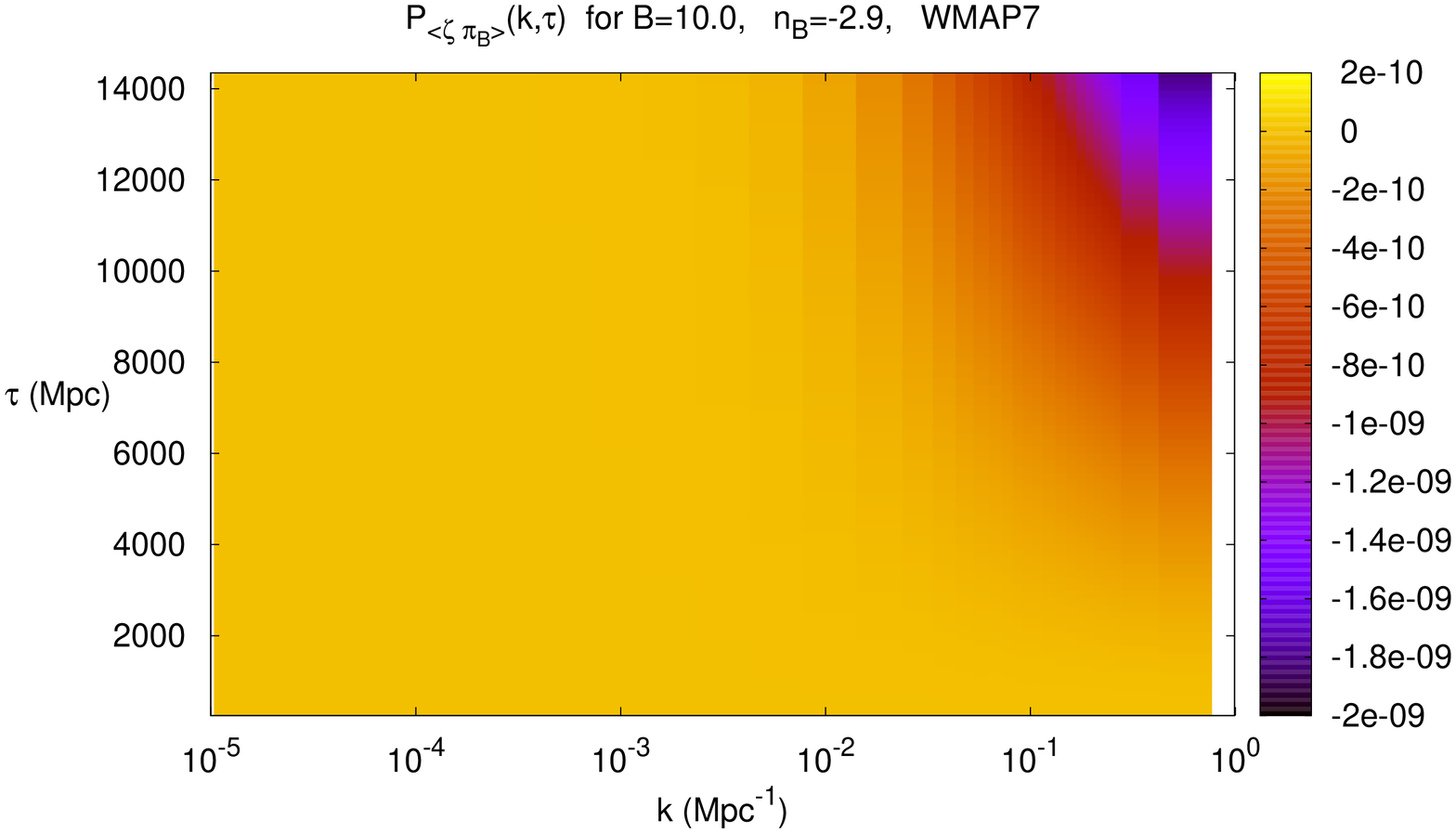}
\hspace{0.2cm}
\epsfxsize=3.2in\epsfbox{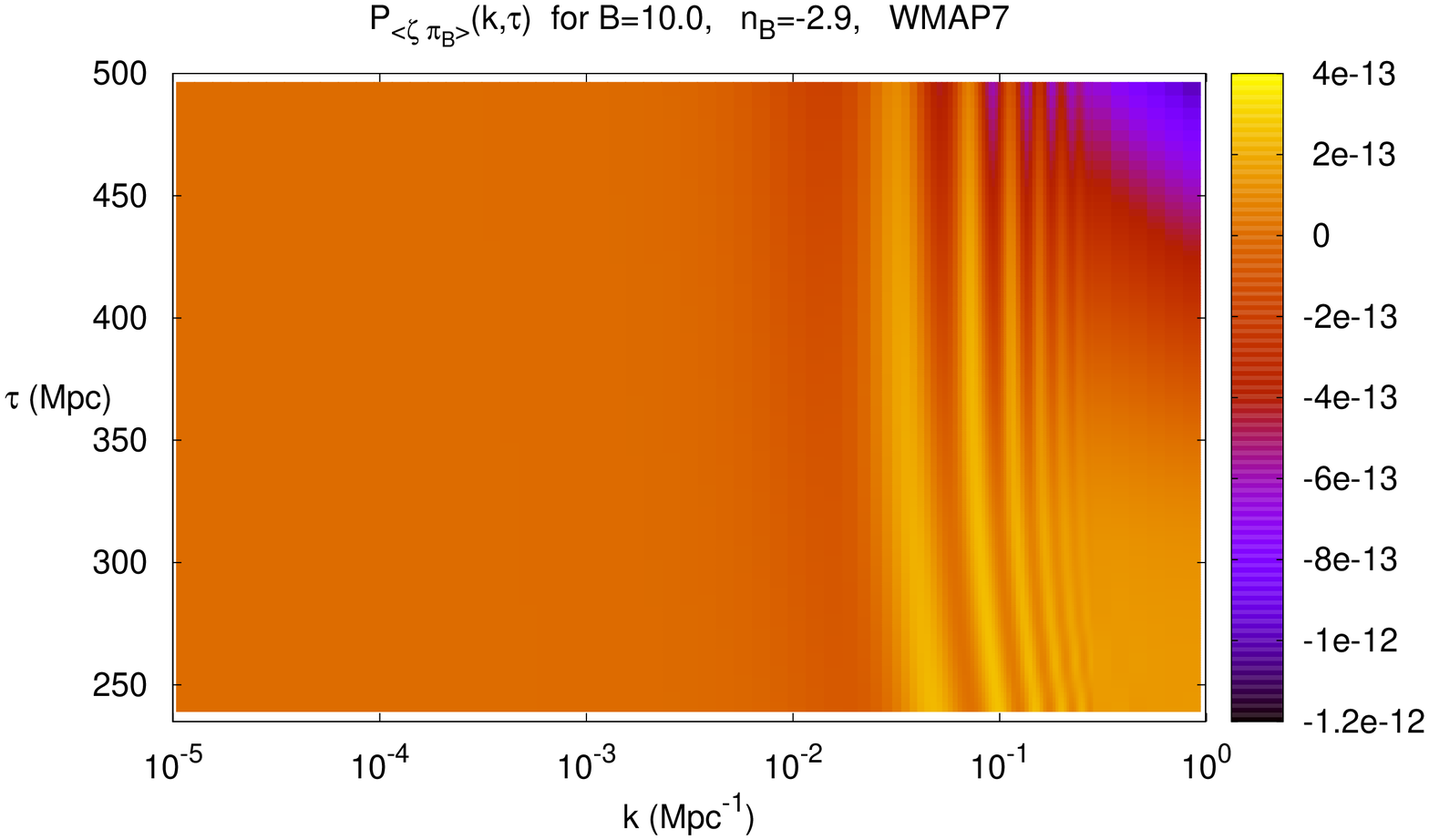}}
\caption{{\it Top panel:} A color map of the values of the dimensionless spectral function determining the cross correlation between the primordial curvature mode and the magnetic energy density contrast  and that of the adiabatic mode ${\cal P}_{\langle\zeta \Delta_B\rangle}(k,\tau)$. 
{\it Bottom panel: } A color map of the values of the dimensionless spectral function determining the cross correlation between the primordial curvature mode and the magnetic anisotropic stress and that of the adiabatic mode ${\cal P}_{\langle\zeta \pi_B\rangle}(k,\tau)$.
In both panels, the figure on the {\it left} shows the whole range of $\tau$ and on the {\it right} the time around last scattering  $\tau_{ls}=285$ Mpc
is shown at a higher resolution. Present time corresponds to $\tau_0=14350$ Mpc.
For the primordial curvature mode the bestfit values of WMAP7 are used. }
\label{fig1}
\end{figure}

In order to estimate the effect on the angular power spectra of the CMB 
the final expressions are obtained using $\Delta_B(\vec{k},\tau)$ and $\pi_B(\vec{k},\tau)$ calculated by employing (\ref{b1}). Moreover,
relevant definitions can be found in the Appendix.
This is in contrast to  previous work where only the amplitude of the magnetic field in $k$-space, $B^{(0)}(\vec{k})$, is taken into account which is constant in time  yielding time independent expressions, 
$\Delta_{B^{(0)}}(\vec{k})$ and $\pi_{B^{(0)}}(\vec{k})$ \cite{cmbmag1}-\cite{kk2}.
As can be seen from (\ref{b1}) the cross correlation between the amplitude of the adiabatic, primordial curvature mode $\zeta(\vec{k})$ and the magnetic energy density contrast $\Delta_B(\vec{k},\tau)$ as well as the anisotropic stress $\pi_B(\vec{k},\tau)$ are non zero. Therefore there  is an additional contribution to the CMB anisotropies.
It is assumed that initially, there is no cross correlation between the primordial curvature mode and the magnetic mode, $\langle\zeta^*(\vec{k}') B_i^{(0)}(\vec{k})\rangle =0$.
At some later time $\tau$ these cross correlations are non vanishing and found to be, at lowest order,
\begin{eqnarray}
\langle \zeta^*(\vec{k}')\Delta_B(\vec{k},\tau)\rangle\simeq \frac{1}{2\rho_{\gamma 0}}\sum_{\vec{q}}\left[
\langle \zeta^*(\vec{k}')b_i(\vec{q},\tau)B^{(0)\; i}(\vec{k}-\vec{q})\rangle+
\langle \zeta^*(\vec{k}')B_i^{(0)}(\vec{q})b^{i}(\vec{k}-\vec{q},\tau)\rangle\right]
\end{eqnarray}
which in the continuum limit together with the expression for the initial magnetic field power spectrum
\begin{eqnarray}
P_{B^{(0)}}(k,k_m)=\frac{2\pi^2}{k^3}\frac{2\rho_B}{\Gamma\left(\frac{n_B+3}{2}\right)}\left(\frac{k}{k_m}\right)^{n_B+3}e^{-\left(\frac{k}{k_m}\right)^2},
\end{eqnarray}
where $\rho_B$ is the magnetic energy density today calculated using the  magnetic field strength smoothed over the magnetic diffusion scale $k_m\simeq 200.694\left(\frac{B}{\rm nG}\right)^{-1}{\rm Mpc}^{-1}$ with a Gaussian window function (cf. \cite{kk1}, \cite{sb}, \cite{jko}) assuming $n_B>-3$, leads to
\begin{eqnarray}
&&\langle\zeta^*(\vec{k}')\Delta_B(\vec{k},\tau)\rangle=\delta_{\vec{k} \vec{k}'}\frac{2\pi^2}{k^3}\frac{A_s}{2\Gamma\left(\frac{n_B+3}{2}\right)}
\frac{\rho_B}{\rho_{\gamma 0}}
\left(\frac{k}{k_m}\right)^{3+n_B}\left(\frac{k}{k_p}\right)^{n_s-1}
\int_0^{\infty}dz z^2e^{-\left(\frac{k}{k_m}\right)^2z^2}
\nonumber\\
&&\times
\int_{-1}^1 dy\left[e^{-\left(\frac{k}{k_m}\right)^2(1-2zy)}\left(1-2zy+z^2\right)^{\frac{n_B-2}{2}}
z\left[-2(1+z^2)y+z(1+3y^2)\right]
\right.
\nonumber\\
&&\left.
\hspace{2cm}+z^{n_B}\left(-1+2zy-y^2\right)\right]F_{V_b^{\zeta}}(x),
\end{eqnarray}
where $y=\vec{k}\cdot\vec{q}/(kq)$, $z\equiv q/k$ and 
\begin{eqnarray}
F_{V_b^{\zeta}}(x)\equiv\int_{x_i}^x dx'V_b^{\zeta}(x')
\end{eqnarray}
for $x\equiv k\tau$ and  $x_i\equiv k\tau_i$.
Similarly, for the cross correlation between the primordial curvature mode and the magnetic anisotropic stress it is found that
\begin{eqnarray}
&&\langle\zeta^*(\vec{k}')\pi_B(\vec{k},\tau)\rangle=\delta_{\vec{k}\vec{k}'}\frac{2\pi^2}{k^3}\frac{3}{2}\frac{A_s}{\Gamma\left(\frac{n_B+3}{2}\right)}\frac{\rho_B}
{\rho_{\gamma 0}}\left(\frac{k}{k_m}\right)^{n_B+3}\left(\frac{k}{k_p}\right)^{n_s-1}\int_0^{\infty}dz z^2 e^{-\left(\frac{k}{k_m}\right)^2z^2}
\nonumber\\
&&\times
\int_{-1}^1 dy\left[e^{-\left(\frac{k}{k_m}\right)^2\left(1-2zy\right)}\left(1-2zy+z^2\right)^{\frac{n_B-2}{2}}
\left(2z^2+(2-z^2)zy-6z^2y^2+3z^3y^3\right)
\right.\nonumber\\
&&\left.
\hspace{2cm}+z^{n_B}\left(1+zy+y^2-3zy^3\right)\right]
F_{V_b^{\zeta}}(x).
\end{eqnarray} 
The  two-point correlation function of a random variable $\chi$ can be written in terms of the dimensionless spectrum ${\cal P}_{\langle \chi\chi\rangle}$
\begin{eqnarray}
\langle \chi_{\vec{k}}^* \chi_{\vec{k}'}\rangle=\frac{2\pi^2}{k^3}{\cal P}_{\langle \chi\chi\rangle}(k)\delta_{\vec{k}\vec{k}'}.
\end{eqnarray}
The dimensionless spectral functions determining the cross correlations between the primordial curvature mode and the magnetic mode , ${\cal P}_{\langle \zeta\Delta_B\rangle}(k,\tau)$ and ${\cal P}_{\langle\zeta\pi_B\rangle}(k,\tau)$,  are shown in figure \ref{fig1} for the best fit values of the 6-parameter $\Lambda$CDM model of WMAP7 \cite{wmap7} at present time, in particular, 
$A_S=2.43\times 10^{-9}$, $n_s=0.963$, $k_p=0.002$ Mpc$^{-1}$, $\Omega_b=0.00227h^{-2}$, $\Omega_{\Lambda}=0.738$, $h=0.714$, and the reionization optical depth $\tau_{re}=0.086$.
Moreover, using a Gaussian window function \cite{kk1}  the smoothed magnetic field strength is set to $B_0=10$ nG and the spectral index $n_B=-2.9$. 
These values are used throughout in all numerical solutions.
The baryon velocity for the primordial curvature mode has been calculated using CMBEASY \cite{cmbeasy}. 
As can be appreciated from figure \ref{fig1} whereas the primordial curvature perturbation is correlated with the 
magnetic energy density ({\it upper panel}) it is anti-correlated with the magnetic anisotropic stress 
({\it lower panel}) in a large part of the $(k, \tau$)-plane.
Moreover, close to decoupling the acoustic oscillations in the baryon-photon fluid manifest themselves in the cross correlations of the 
curvature mode and the magnetic mode. Visually this results in the "ripples" in the color maps ({\it right-hand-side column}).
The amplitudes of the dimensionless spectral functions determining the auto and cross correlation functions of constant $\Delta_{B^{(0)}}$ and $\pi_{B^{(0)}}$ for $B_0=10$ nG and $n_B=-2.9$ for the same range of $k$ values are of the order of $10^{-11}$ for $\langle\pi_{B^{(0)}}^*\pi_{B^{(0)}}\rangle$ as well as 
$\langle\Delta_{B^{(0)}}^*\pi_{B^{(0)}}\rangle$ 
and of the order of $10^{-12}$ for $\langle\Delta_{B^{(0)}}^*\Delta_{B^{(0)}}\rangle$ \cite{kk1}. 
Thus comparing these zeroth order auto and cross correlation functions with the first order cross correlation functions in figure \ref{fig1} ({\it right-hand-side}) around decoupling shows that the latter are significantly lower than the former.

In \cite{cmk} a non trivial cross-correlation between the primordial curvature mode and the magnetic field resulted during inflation by assuming 
that the magnetic field is generated during inflation and conformal invariance is broken by coupling the electromagnetic field to the curvaton field.
This is different from the case considered here where the non vanishing cross correlations of the magnetic field and the primordial curvature perturbation are due to the backreaction of the baryon fluid onto the magnetic field.

Finally, estimating $\langle\Delta^*_B(\vec{k},\tau)\Delta_B(\vec{k}',\tau)\rangle$, $\langle\Delta^*_B(\vec{k},\tau)\pi_B(\vec{k}',\tau)\rangle$
and $\langle\pi^*_B(\vec{k},\tau)\pi_B(\vec{k}',\tau)\rangle$ shows that the first order corrections to the zeroth order expressions are of the order of 
$\left[P_{B^{(0)}}(k,k_m)\right]^3$ whereas the zeroth order expressions are of the order of $\left[P_{B^{(0)}}(k,k_m)\right]^2$. Taking into account that the magnetic field spectrum is nearly scale-invariant $n_B=-2.9$ these corrections are suppressed, with respect to the zeroth order expressions,  by a factor $\frac{\rho_B}{\rho_{\gamma0}}\simeq 10^{-7}\left(\frac{B}{\rm nG}\right)^2$  \cite{kk1}, which for the values at hand is of the order of $10^{-5}$, and will therefore be neglected. 
This is the same factor which controls the corrections to the spectral function determining the two point function of the magnetic field.

\subsection{Vector modes}

Before closing this section we briefly comment on vector modes. In this case the perturbation of the fluid 3-velocity is determined by the gauge-invariant amplitude of the  matter velocity field $V^{(\pm 1)}(\vec{k},\tau)$ \cite{ks}, so that the induction equation (\ref{e2}), implies in terms of the helicity basis, defined in the Appendix (cf. equation  (\ref{heli}))
\begin{eqnarray}
\dot{B}_i(\vec{k},\tau)=\mp i\sum_{\vec{q}}\left[\left(\hat{e}^{\pm}_{\vec{q}}\right)_iq_j-\left(\hat{e}^{\pm}_{\vec{q}}\right)_mk_m\delta_{ij}\right]B_j(\vec{k}-\vec{q},\tau)V_b^{(\pm 1)}(\vec{q},\tau).
\end{eqnarray}
 Then for $B_i(\vec{k},\tau)=B_i^{(0)}(\vec{k})+b_i(\vec{k},\tau)$, where $|b_i(\vec{k},\tau)|\ll |B_i^{(0)}(\vec{k})|$ the first order solution is given by
 \begin{eqnarray}
 b_i(\vec{k},\tau)=\mp i \sum_{\vec{q}}\left[\left(\hat{e}_{\vec{q}}^{\pm}\right)_i q_j-\left(\hat{e}_{\vec{q}}^{\pm}\right)_mk_m\delta_{ij}\right]B_j^{(0)}(\vec{k}-\vec{q}) \; \pi_{B^{(0)}}^{(\pm 1)}\left(\vec{q}\right)\int_{\tau_i}^{\tau}d\tau'V_b^{\pi_{B^{(0)}}^{(\pm 1)}}(q,\tau')
 \end{eqnarray}
 where for the baryon velocity a similar notation to equation (\ref{vel}) was used.
The CMB anisotropies are determined by the two-point function to first order 
\begin{eqnarray}
\langle\pi_B^{(+1)*}(\vec{k},\tau)\pi_B^{(+1)}(\vec{k}',\tau)+\pi_B^{(-1)*}(\vec{k},\tau)\pi_B^{(-1)}(\vec{k}',\tau)\rangle&\simeq&
\langle\pi_{B^{(0)}}^{(+1)*}(\vec{k})\pi_{B^{(0)}}^{(+1)}(\vec{k}')+\pi_{B^{(0)}}^{(-1)*}(\vec{k})\pi_{B^{(0)}}^{(-1)}(\vec{k}')\rangle
\nonumber\\
+
\langle\pi_{B^{(0)}}^{(+1)*}(\vec{k})\pi_B^{(+1)}(\vec{k}',\tau)+\pi_{B^{(0)}}^{(-1)*}(\vec{k})\pi_B^{(-1)}(\vec{k}',\tau)\rangle
&+& (\vec{k}\leftrightarrow\vec{k}')
\end{eqnarray}
The scaling with the number of powers of the magnetic field spectrum $P_{B^{(0)}}(k,k_m)$ is the same as
in the scalar case. Therfore the corrections to the zeroth order expression are suppressed by  a factor $\frac{\rho_B}{\rho_{\gamma0}}$ \cite{kk2}, which for the values at hand is of the order of $10^{-5}$, and will therefore be neglected.

\section{Resulting angular power spectra}

\begin{figure}[h!]
\centerline{\epsfxsize=3.1in\epsfbox{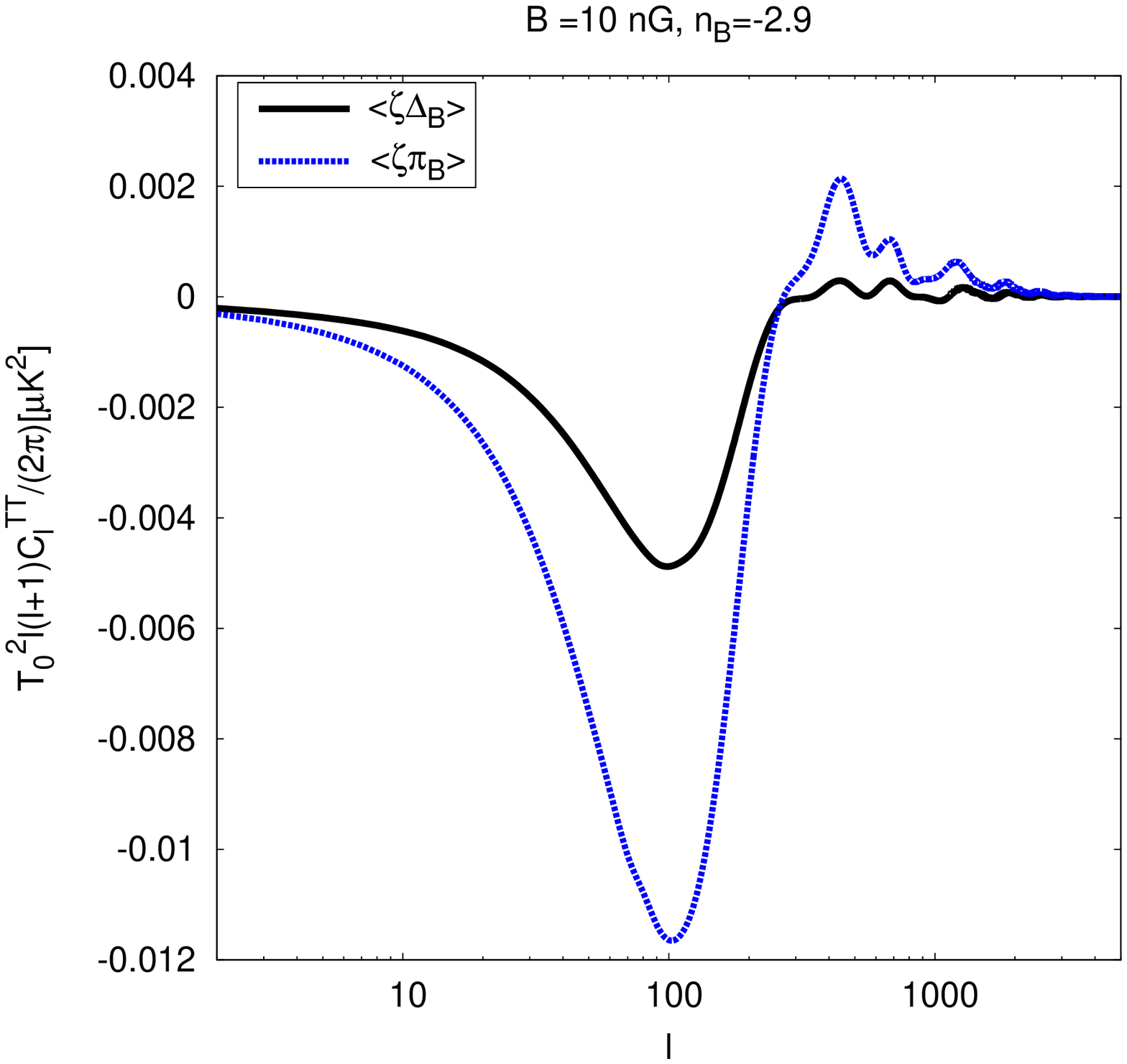}
\hspace{0.2cm}
\epsfxsize=3.1in\epsfbox{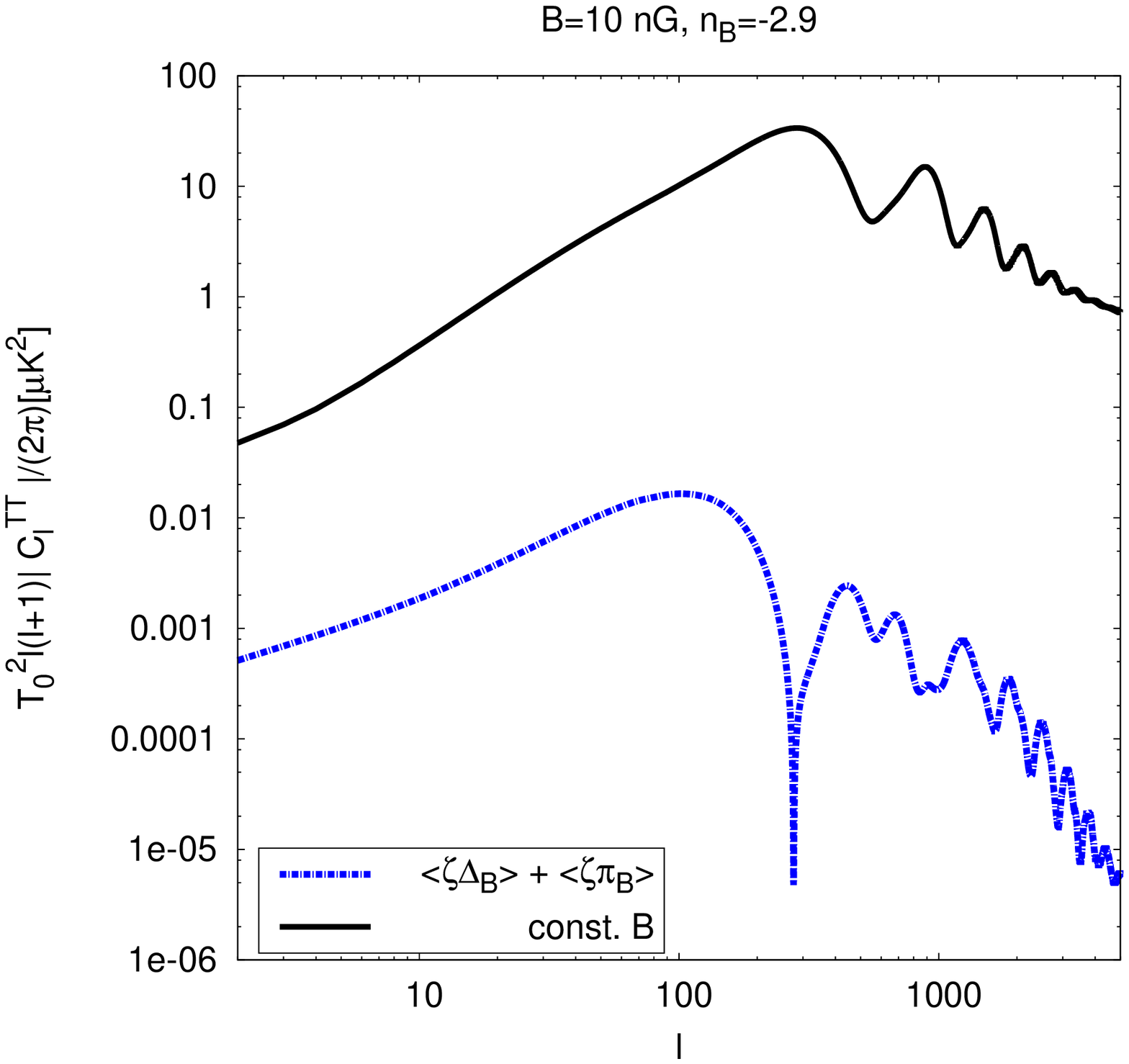}}
\caption{The angular spectrum determining the temperature autocorrelation function $C_{\ell}^{\rm TT}$. {\it Left:} $C_{\ell}^{TT,\langle \zeta\Xi\rangle}$ due to the cross correlation between the primordial curvature mode and the magnetic mode. {\it Right:} The total angular power spectrum due to the correlated magnetic curvature mode, $C_{\ell}^{TT,\langle \zeta\Delta_B\rangle} + C_{\ell}^{TT,\langle \zeta\pi_B\rangle} $
 in comparison with that due to a constant magnetic field, 
 $C_{\ell}^{TT,\langle\Delta_{B^{(0)}}\Delta_{B^{(0)}}\rangle} + 
2 C_{\ell}^{TT,\langle \Delta_{B^{(0)}}\pi_{B^{(0)}}\rangle}+
 C_{\ell}^{TT,\langle \pi_{B^{(0)}}\pi_{B^{(0)}}\rangle}$
 \cite{kk1}.} 
\label{fig2}
\end{figure}

\begin{figure}[h!]
\centerline{\epsfxsize=3.1in\epsfbox{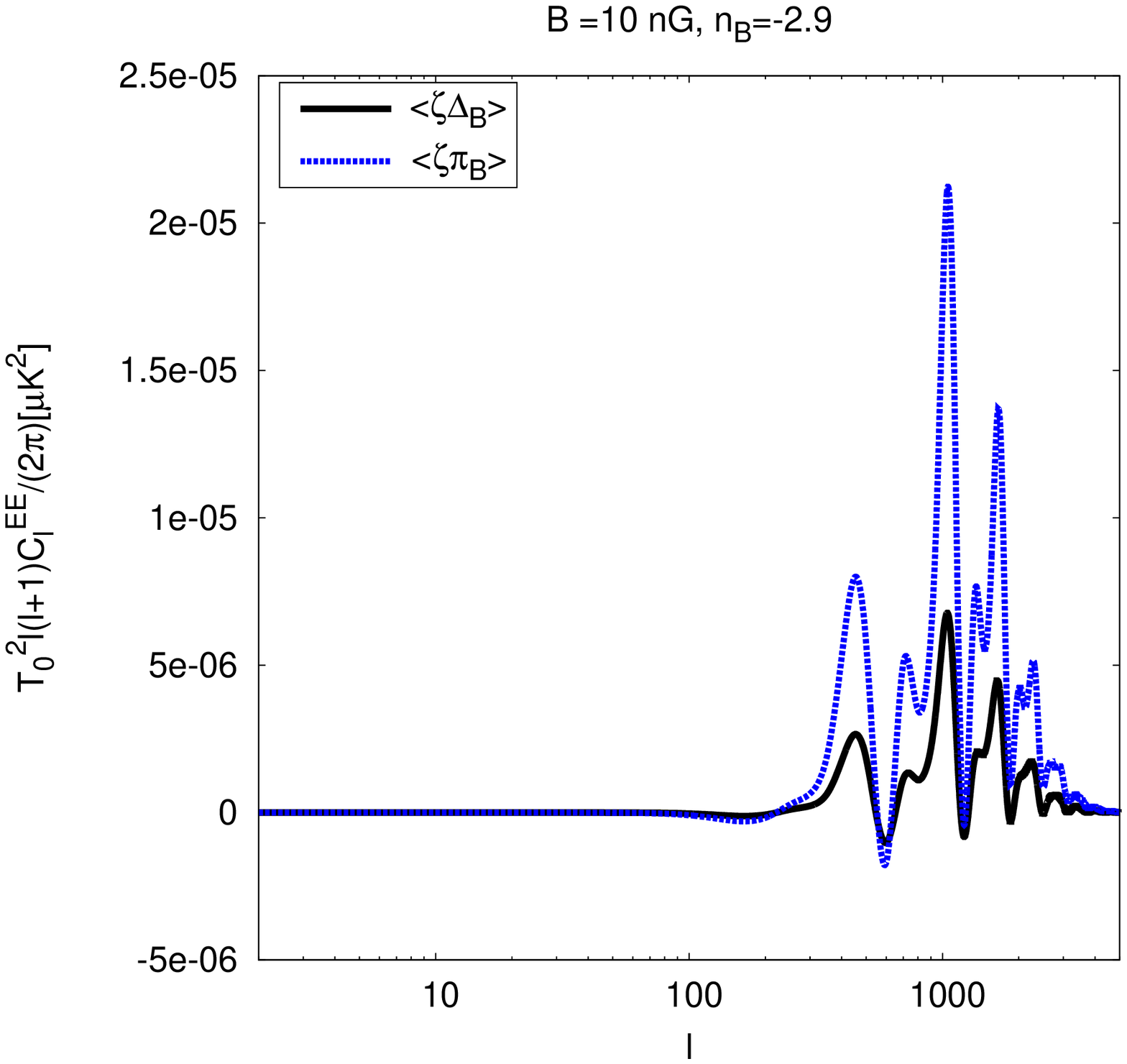}
\hspace{0.2cm}
\epsfxsize=3.1in\epsfbox{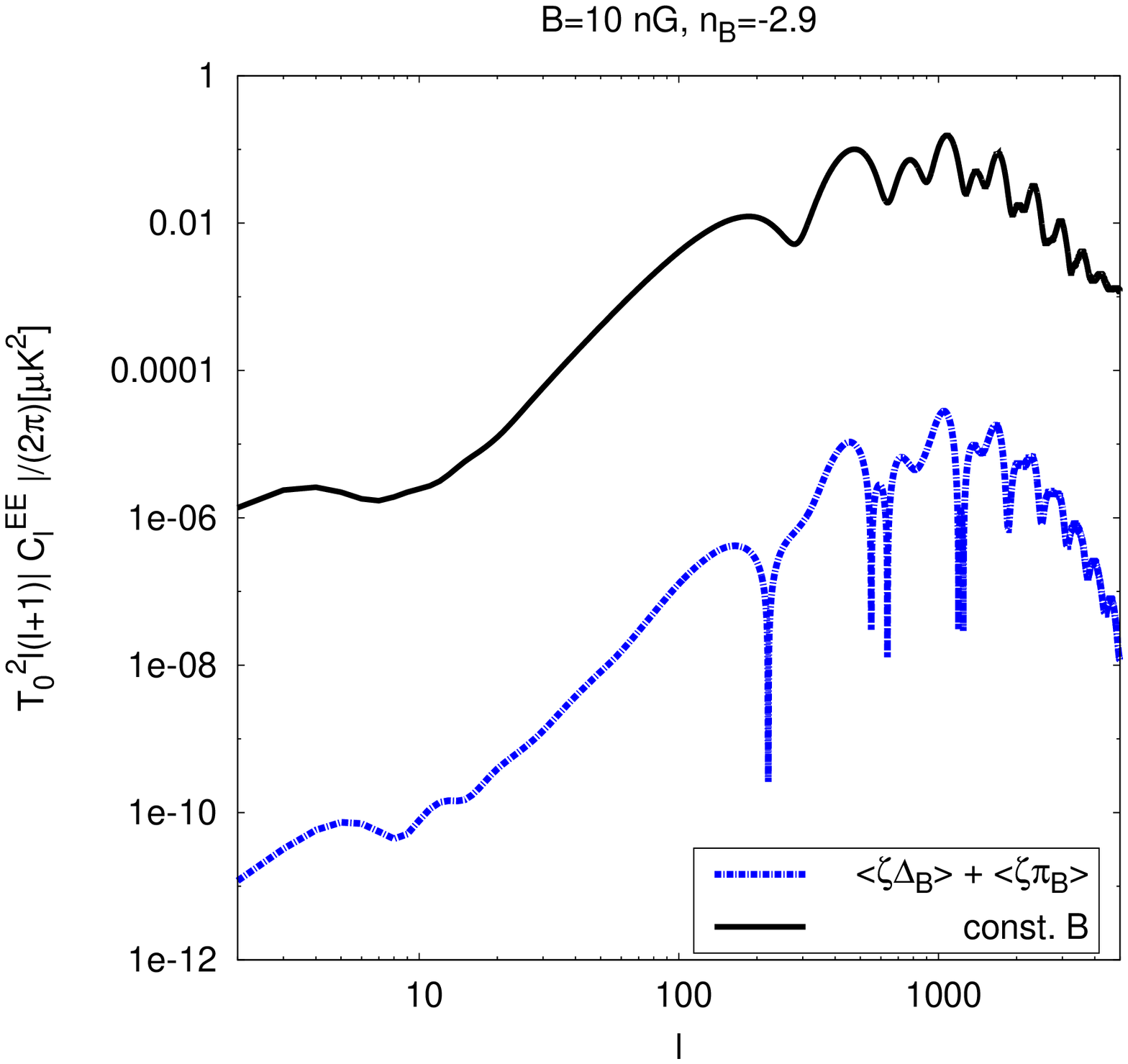}}
\caption{The angular spectrum determining the polarization E-mode autocorrelation function $C_{\ell}^{\rm EE}$. {\it Left:} $C_{\ell}^{EE,\langle \zeta\Xi\rangle}$ due to the cross correlation between the primordial curvature mode and the magnetic mode. {\it Right:} The total angular power spectrum due to the correlated magnetic curvature mode, $C_{\ell}^{EE,\langle \zeta\Delta_B\rangle} + C_{\ell}^{EE,\langle \zeta\pi_B\rangle} $
 in comparison with that due to a constant magnetic field, 
 $C_{\ell}^{EE,\langle\Delta_{B^{(0)}}\Delta_{B^{(0)}}\rangle} + 
2 C_{\ell}^{EE,\langle \Delta_{B^{(0)}}\pi_{B^{(0)}}\rangle}+
 C_{\ell}^{EE,\langle \pi_{B^{(0)}}\pi_{B^{(0)}}\rangle}$
 \cite{kk1}.} 
\label{fig3}
\end{figure}

\begin{figure}[h!]
\centerline{\epsfxsize=3.1in\epsfbox{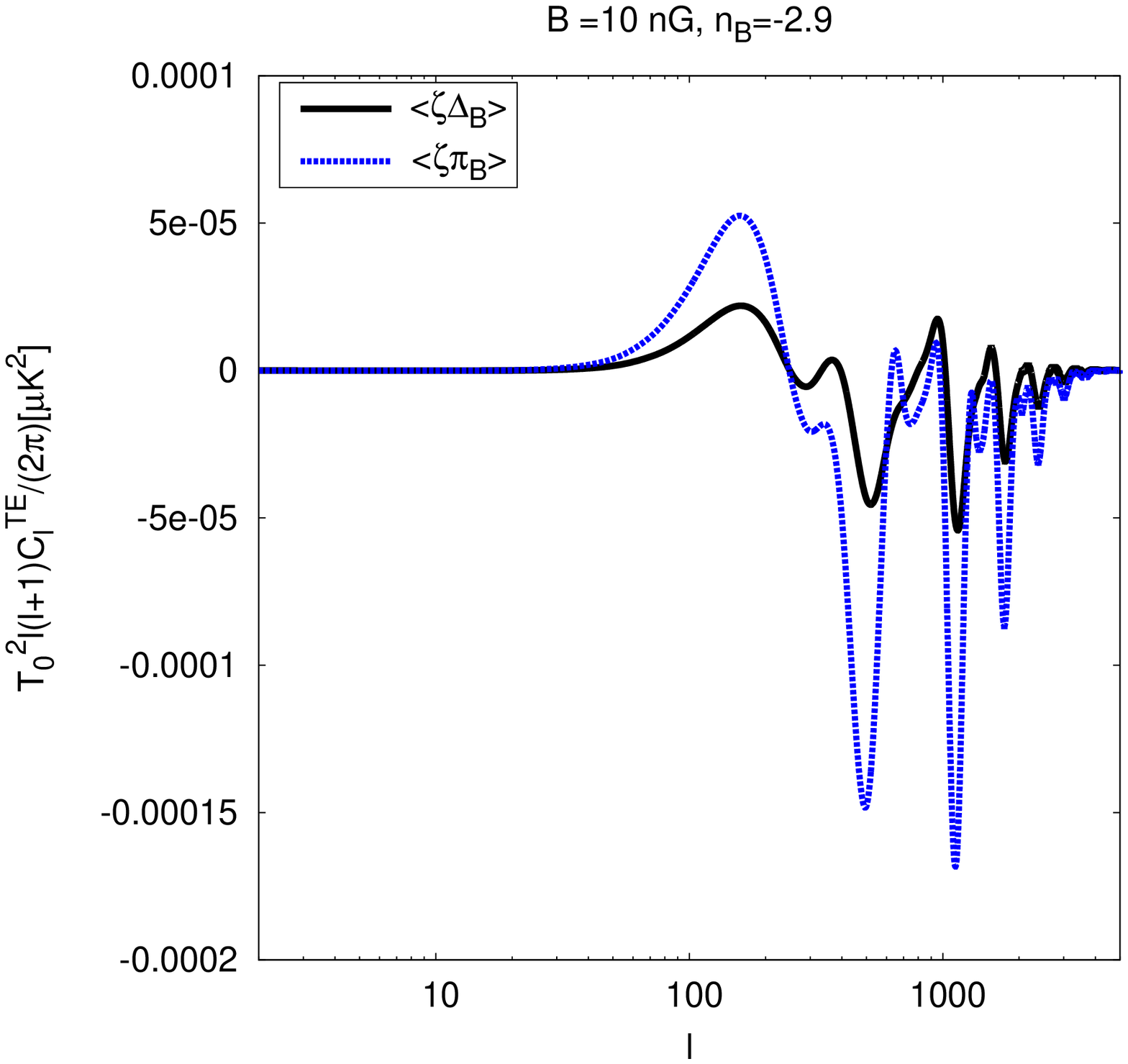}
\hspace{0.2cm}
\epsfxsize=3.1in\epsfbox{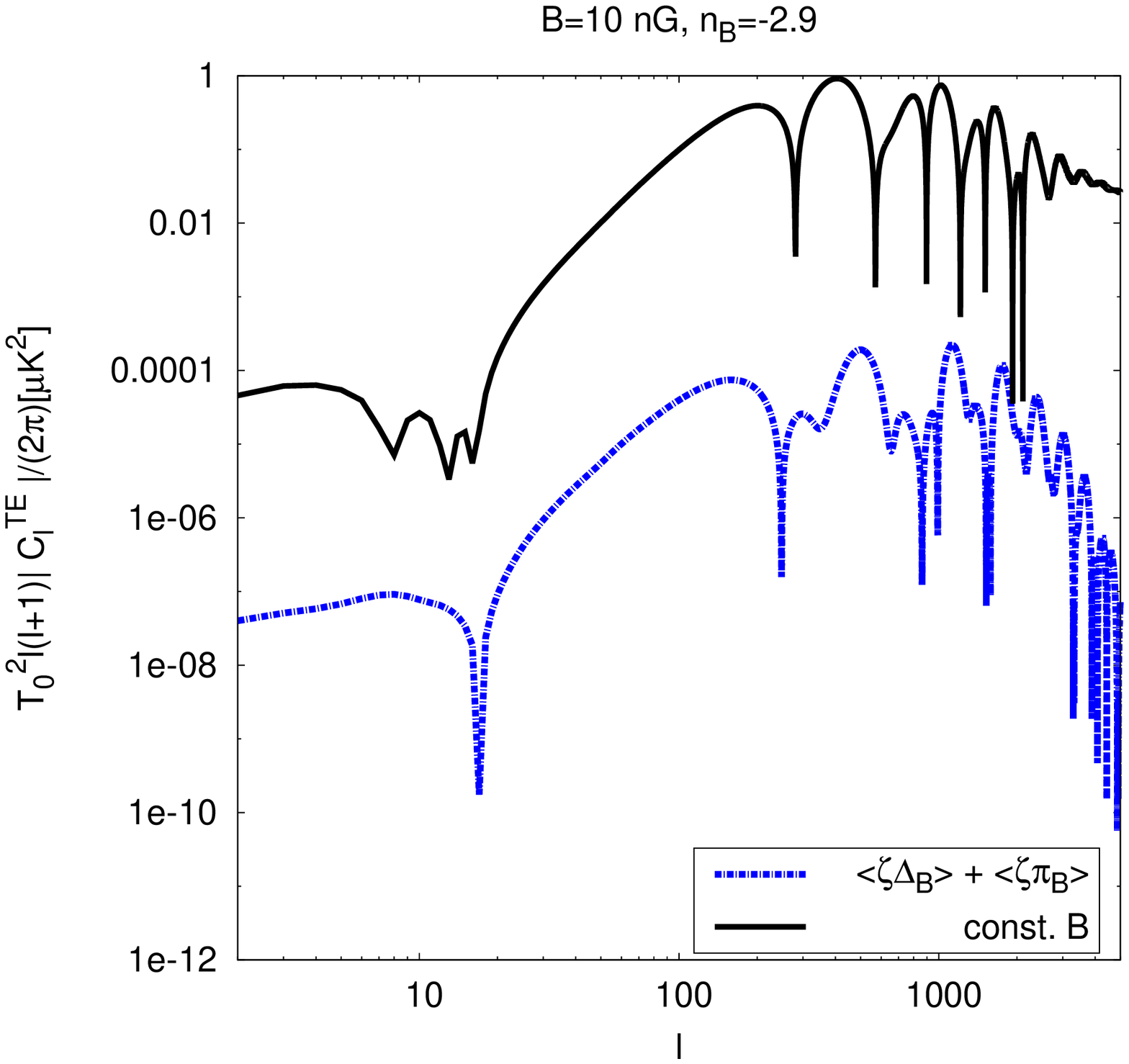}}
\caption{The angular spectrum determining the temperature polarization cross correlation function $C_{\ell}^{\rm TE}$. {\it Left:} $C_{\ell}^{TE,\langle \zeta\Xi\rangle}$ due to the cross correlation between the primordial curvature mode and the magnetic mode. {\it Right:} The total angular power spectrum due to the correlated magnetic curvature mode, $C_{\ell}^{TE,\langle \zeta\Delta_B\rangle} + C_{\ell}^{TE,\langle \zeta\pi_B\rangle} $
 in comparison with that due to a constant magnetic field, 
 $C_{\ell}^{TE,\langle\Delta_{B^{(0)}}\Delta_{B^{(0)}}\rangle} + 
2 C_{\ell}^{TE,\langle \Delta_{B^{(0)}}\pi_{B^{(0)}}\rangle}+
 C_{\ell}^{TE,\langle \pi_{B^{(0)}}\pi_{B^{(0)}}\rangle}$
 \cite{kk1}.} 
\label{fig4}
\end{figure}

The line-of-sight integration approach \cite{sz}  allows to write the brightness function for each component as \cite{hw}, for the scalar mode,
\begin{eqnarray}
\frac{\Theta^X_{\ell}(\tau_0,k)}{2\ell +1}=\int_0^{\tau_0} d\tau S_{\Theta}^X(k,\tau)j_{\ell}\left[k(\tau_0-\tau)\right]
\end{eqnarray}
where $S_{\Theta}^X$ is the source function and $X$ denotes $\zeta$, $\Delta_B$ and $\pi_B$.
Thus the angular power spectra determining the temperature autocorrelation function are given by
\begin{eqnarray}
C_{\ell}^{TT,\langle \zeta\Xi\rangle}=\frac{1}{2\pi^2}\int_0^{\infty}\frac{dk}{k}\int_0^{\tau_0}d\tau{\cal P}_{\langle \zeta\Xi\rangle}(k,\tau)S_{\Theta}^{\zeta}(k,\tau)S_{\Theta}^{\Xi}(k,\tau)
\left(j_{\ell}\left[k(\tau_0-\tau)\right]\right)^2
\end{eqnarray}
where $\Xi=\Delta_B, \pi_B$
and a similar expression for the autocorrelation functions of the E-mode. The cross-correlation between temperature ($T$) and polarization ($E$) is determined accordingly by
\begin{eqnarray}
C_{\ell}^{TE,\langle \zeta\Xi\rangle}&=&\frac{1}{2\pi^2}\int\frac{dk}{k}\int d\tau \left[
{\cal P}_{\langle \zeta\Xi\rangle}(k,\tau)S_{\Theta}^{\zeta}(k,\tau)S_{E}^{\Xi}(k,\tau)
\right.\nonumber\\
&&\hspace{3cm}\left.+
{\cal P}_{\langle \zeta\Xi\rangle}(k,\tau)S_{E}^{\zeta}(k,\tau)S_{\Theta}^{\Xi}(k,\tau)
\right]\left(j_{\ell}\left[k(\tau_0-\tau)\right]\right)^2.
\end{eqnarray}
The resulting angular power spectra are calculated in a modified version of the numerical code of \cite{kk1} which is based on CMBEASY \cite{cmbeasy}.
The results for the bestfit values WMAP7 and magnetic field as specified before are shown in figures \ref{fig2}-\ref{fig4}. 
The presence of a magnetic field before neutrino decoupling results in a source term for the evolution equation of the amplitude of the comoving 
curvature perturbation. Therefore there is a contribution in addition to the primordial curvature perturbation generated during inflation.
After neutrino decoupling the comoving curvature perturbation is constant on superhorizon scales. Moreover, the contributions due to the neutrino 
anisotropic stress and the magnetic anisotropic stress cancel invoking the compensated mode \cite{ls}.
The initial conditions for the numerical solutions are set after neutrino decoupling. Hence the total comoving curvature perturbation is given by
\begin{eqnarray}
\zeta^{total}_{\vec{k}}=\zeta_{\vec{k}}+\zeta_{\vec{k}}^{B^{(0)}}
\end{eqnarray}
where $\zeta_{\vec{k}}^{B^{(0)}}=-\frac{\Omega_{\nu}\pi_{B^{(0)}}}
{3\left(\Omega_{\gamma}+\Omega_{\nu}\right)}\beta$ with $\beta\equiv\ln\frac{\tau_{\nu}}{\tau_{B}}$, and  $\tau_B$ and $\tau_{\nu}$ being the time of (instantaneous) generation of the magnetic field and the time of neutrino decoupling, respectively \cite{ls}. However, as the additional contribution to the curvature perturbation is proportional to the amplitude of the magnetic anisotropic stress it can be neglected in the calculation of the cross correlation between the curvature perturbation and the magnetic field variables.
Therefore, the comparison with the mode due to a constant magnetic field in $k$-space, $B_i^{(0)}(\vec{k})$,  in figures \ref{fig2}-\ref{fig4} is done 
using the numerical solution of the compensated mode.

\section{Conclusions}

The evolution of a primordial magnetic field present since long before decoupling has been calculated approximately for scalar and vector modes by solving the induction equation by iteration to first order. This effective backreaction of the baryon velocity on the magnetic field causes a non trivial time evolution of the magnetic field in Fourier space in contrast with the constant time behaviour at zeroth order corresponding in real space to the redshifting of the magnetic field strength with the expansion of the universe. 
Observations of the CMB clearly indicate the presence of an adiabatic mode which could be a primordial curvature perturbation generated during inflation 
\cite{wmap7}. 
Therefore, considering a stochastic, gaussian magnetic field, taking into account the effect of the baryon velocity induces non 
vanishing cross correlations between the primordial curvature mode generated during inflation and the scalar magnetic mode. 
This is what would be expected by the nonlinear interaction of the magnetic field with the baryon fluid. 
Using the zeroth order numerical solutions for the source functions in the line of sight integration in addition to the spectral functions determining the cross correlation functions between the primordial curvature modes and the magnetic mode the angular power spectra determining the temperature anisotropies and 
polarization of the CMB are calculated numerically. These are the first order corrections due to an evolving magnetic field in Fourier space. The numerical results show that these corrections are several orders below the zeroth order results and can therefore be neglected, e.g., in the estimation of parameters including a primordial magnetic field using the angular power spectra of the CMB \cite{par-est}.
The first order corrections to the auto and cross correlation functions of the magnetic energy density and the magnetic anisotropic stress of the scalar magnetic mode have been estimated to be much below that of the zeroth order expressions and therefore their imprint on the CMB angular power spectra will be even smaller. The same conclusion holds for the first order correction to the zeroth order correlation function of the magnetic anisotropic stress of the vector mode.

\section{Acknowledgements}

I am indebted to Angela Olinto for interesting discussions and comments on the manuscript.
Financial support by Spanish Science Ministry grants FPA2009-10612, FIS2009-07238 and CSD2007-00042 is gratefully acknowledged.

\section{Appendix}
\label{app}
\setcounter{equation}{0}

The energy density of the magnetic field  is written in terms of the gauge invariant magnetic energy contrast $\Delta_B$ such that
\begin{eqnarray}
\rho_B=\rho_{\gamma}\sum_{\vec{k}}\Delta_B(\vec{k},\tau)Q^{(0)}(\vec{k},\vec{x}),
\end{eqnarray}
where $Q^{(0)}(\vec{k},\vec{x})$ denote a set of scalar harmonic functions satisfying $(\Delta+k^2)Q^{(0)}=0$
(cf.  e.g. \cite{ks}).
The magnetic anisotropic stress is determined by
\begin{eqnarray}
\pi_{(ij)}(\vec{x},\tau)=p_{\gamma}\sum_{m=0,\pm1, \pm 2}\sum_{\vec{k}}\pi_B^{(m)}(\vec{k},\tau)Q_{ij}^{(m)}(\vec{k},\vec{x}),
\label{pij}
\end{eqnarray}
where $m=0$ denotes the scalar part and $Q_{ij}^{(0)}=k^{-2}Q_{|ij}+\frac{1}{3}Q^{(0)}$,
the vector part is determined by $m=\pm 1$ and $Q_{ij}^{(\pm 1)}=-\frac{1}{2k}\left(Q_{i|j}^{(\pm 1)}+Q_{j|i}^{(\pm 1)}\right)$ and the tensor modes are given by $m=\pm 2$ \cite{hw}. However, here only the scalar and vector modes will be relevant.
Expanding the magnetic field as
$
a^2 B_i(\vec{x},\tau)=a_0^2\sum_{\vec{k}}B_i(\vec{k},\tau)Q^{(0)}(\vec{k},\vec{x})
$
yields to
\begin{eqnarray}
\Delta_B(\vec{k},\tau)=\frac{1}{2\rho_{\gamma 0}}\sum_{\vec{q}}B_i(\vec{q},\tau)B^i(\vec{k}-\vec{q},\tau).
\end{eqnarray}
A convenient representation of the scalar and vector harmonic 
functions in flat space is given by \cite{Thorne,Rose,hw}
\begin{eqnarray}
Q^{(0)}(\vec{k},\vec{x})&=&e^{i\vec{k}\cdot\vec{x}}\\
Q^{(\pm 1)}(\vec{k},\vec{x})_i&=&\pm\frac{i}{\sqrt{2}}\left(\hat{e}_1\pm i\hat{e}_2\right)_ie^{i\vec{k}\cdot\vec{x}}
\label{A1}
\end{eqnarray}
The (spatial) coordinate system defined by the unit vectors $\hat{e}_1$, $\hat{e}_2$ and $\hat{e}_3$ is chosen such that $\hat{e}_3$ lies in the direction of $\vec{k}$, thus  $\hat{e}_3=\hat{k}$.
Moreover, in  the helicity basis \cite{cdk}
\begin{eqnarray}
\hat{e}^{\pm}_{\vec{k}}=-\frac{i}{\sqrt{2}}\left(\hat{e}_1\pm i\hat{e}_2\right)
\label{heli}
\end{eqnarray}
so that $\hat{e}_{\vec{k}}^\pm\cdot\hat{e}_{\vec{k}}^{\mp}=-1$ and $\hat{e}^{\pm}_{\vec{k}}\cdot\hat{e}_{\vec{k}}^{\pm}=0$ and  $\hat{e}^{\pm}_{\vec{k}}\cdot\hat{k}=0$.
With this choice the scalar and vector parts of the anisotropic stress are found to be (e.g. \cite{kk2})
\begin{eqnarray}
\pi_B^{(0)}(\vec{k},\tau)&=&\frac{3}{2\rho_{\gamma0}}\left[\sum_{\vec{q}}\frac{3}{k^2}B_i(\vec{k}-\vec{q},\tau)q^iB_j(\vec{q},\tau)\left(k^j-q^j\right)-\sum_{\vec{q}}B_m(\vec{k}-\vec{q},\tau)B^m(\vec{q},\tau)\right]
\label{p0}\\
\pi_B^{(\pm 1)}(\vec{k},\tau)&=&\mp i\frac{3}{\rho_{\gamma 0}}\sum_{\vec{q}}\left[\left(\hat{e}_{\vec{k}}^{\mp}\right)^iB_i(\vec{k}-\vec{q},\tau)B_j(\vec{q},\tau)\hat{k}^j+\left(\hat{e}_{\vec{k}}^{\mp}\right)^jB_j(\vec{q},\tau)B_i(\vec{k}-\vec{q},\tau)\hat{k}^i\right].
\label{p2}
\end{eqnarray}
For simplicity, $\pi_B^{(0)}$ is denoted as $\pi_B$.

\end{document}